\documentclass[12pt]{article}
\usepackage{amssymb}

\begin{document}

%---------------------Greek letters-------------------------------
\def\a{\alpha}   \def\b{\beta}    \def\c{\chi}       \def\d{\delta}
\def\D{\Delta}   \def\e{\epsilon} \def\f{\phi}       \def\F{\Phi}
\def\g{\gamma}   \def\G{\Gamma}   \def\k{\kappa}     \def\l{\lambda}
\def\L{\Lambda}  \def\m{\mu}      \def\n{\nu}        \def\r{\varrho}
\def\o{\omega}   \def\O{\Omega}   \def\p{\psi}       \def\P{\Psi}
\def\s{\sigma}   \def\S{\Sigma}   \def\vt{\vartheta} \def\t{\tau}
\def\vf{\varphi} \def\x{\xi}      \def\z{\zeta}      \def\th{\theta}
\def\Th{\Theta}
\def\ve{\varepsilon}

%---------------------Caliber letters-------------------------------
\def\CAG{{\cal A/\cal G}}
\def\CA{{\cal A}} \def\CB{{\cal B}}  \def\CC{{\cal C}}
\def\CD{{\cal D}} \def\CF{{\cal F}}  \def\CG{{\cal G}}
\def\CH{{\cal H}} \def\CI{{\cal I}}  \def\CL{{\cal L}}
\def\CO{{\cal O}} \def\CP{{\cal P}}  \def\CS{{\cal S}}
\def\CR{{\cal R}} \def\CT{{\cal T}}  \def\CU{{\cal U}}
\def\CW{{\cal W}} \def\CM{{\cal M}}  \def\CX{{\cal X}}
%%%%%%%%%%%%%%%%%%%%%%%%%%%%%%%%%%%%%%%%%%%%%%%%%%%%%%%%%%%%%%%%%%
\newcommand{\be}{\begin{equation}}
\newcommand{\bea}{\begin{eqnarray}}
\newcommand{\ee}{\end{equation}}
\newcommand{\eea}{\end{eqnarray}}

\begin{titlepage}

%---------------- preprint number & date ---------------
\hfill\parbox{4cm} {KIAS-P00056 \\ KEK-TH-708 \\ hep-th/0008213 \\
August 2000}

%------------------------ title ------------------------
\vspace{15mm}
\begin{center}
{\Large \bf
%Matrix Theory and OM Theory\\
%or\\
U-duality Between NCOS Theory and Matrix Theory}
\end{center}

%---------------- authors and addresses ----------------
\vspace{5mm}
\begin{center}
Seungjoon Hyun\footnote{\tt hyun@kias.re.kr}
\\[5mm]
{\it
Theory Group, KEK, Tsukuba, Ibaraki 305-0801, Japan \\
and \\
School of Physics, Korea Institute for Advanced Study,
Seoul 130-012, Korea \\
}
\end{center}
\thispagestyle{empty}

%----------------------- abstract ----------------------
\vfill
\begin{center}
{\bf Abstract}
\end{center}
\noindent We show that the NCOS (noncommutative open string)
theories on torus $T^p$ ($p\leq 5$) are U-dual to matrix theory on
torus with electric flux background. Under U-duality, the number
of D-branes and the number of units of electric flux get
interchanged. Furthermore under the same U-duality the decoupling
limit taken in the NCOS theory maps to the decoupling limit taken
in the matrix theory, thus ensure the U-duality between those two
class of theories. We consider the energy needed for Higgsing
process and some bound states with finite energy and find
agreements in both theories. \vspace{2cm}
\end{titlepage}

%-------------------------------- Begin ------------------------------

\baselineskip 7mm
\renewcommand{\thefootnote}{\arabic{footnote}}
\setcounter{footnote}{0}
\section{Introduction}

% introduce the NCOS theory
Recently there have been a lot of interests in the NCOS
(noncommutative open string) theory \cite{SST,GMMS,GRS,BR}.  The
theory is realized as the world volume theory of D-branes with
constant electric field which are taken to the critical value. To
extract the worldvolume degrees of freedom only, we need to take
some special limit \cite{SST,GMMS} on the background under which
all the closed string modes in bulk are decoupled. In contrast to
other well-known cases, the higher massive modes of open strings
are not decoupled in the limit and thus the theory is not just
Yang-Mills theory, but the full open string theory defined on the
noncommutative spacetime due to the electric field on the branes.
In this way it becomes a consistent theory involving open strings
only. In \cite{GMSS} whole class of the NCOS theories, OM (open
membrane) theory and OD$p$ (open D$p$-brane) theories are defined.
See also the independent work on OM theory in \cite{BBSS} and the
independent work on OD1 and OD2 theories in \cite{Harm2} with the
name (1,1) OBLST and (2,0) OBLST, respectively. All these theories
are in the same moduli space if they are defined on torus.

One of the most obvious thing to do toward the understanding of
the theory is the study of the duality of the theory, inherited
from the `mother' M/string theory. In the case of (3+1) dimensions
the NCOS theory with $F_{01}\neq 0$ naturally arises
\cite{GMMS,GRS} as the S-dual of the NCYM (noncommutative
Yang-Mills) theory
 with $B_{23}\neq 0$ considered in \cite{SW}. (2+1)-dimensional NCOS is
S-dual to (2+1)-dimensional U(1) Yang-Mills theory on a  M2-brane
at low energy \cite{GMSS}. (1+1)-dimensional NCOS theory is S-dual
\cite{GMSS,GKP,kle} to $(n,1)$-string theory  from which many
insightful observations made in \cite{KM}. See also
\cite{BBSS,KT}. For the related issues and more recent
developments in NCOS theories, see
\cite{GM,CW,Harm,RSJ,Rey,Rey2,Cai,GK,LRS,AOR}.

In this paper we study the NCOS theory on torus. In particular we
show that there are U dualities between the class of matrix theory
\cite{DKPS,BFSS,susskind,seiberg,sen} defined on torus and the
class of the NCOS theory on torus. On one hand it is described by
the large $N$ limit of $U(N)$ SYM (super Yang-Mills) theory or
their completions, in the sector of $M$ units of electric flux due
to the winding fundamental strings.\footnote{As is well-known,
matrix theory on $T^p$ is described by $U(N)$ SYM (super
Yang-Mills) theory on $T^p$ for $p\leq 3$. On higher-dimensional
torus, it is true only as the low energy effective description.
For complete descriptions, we need extra ingredients
\cite{bound,strom,LST}.} On the other hand it is described by the
NCOS theory of $M$ D$p$-branes with background $N$ electric flux
in the limit of critical value. This is deeply related to the
duality between the theory of $N$ D$p$-branes with $M$ fundamental
strings and the theory of $M$ D$p$-branes  with $N$ fundamental
strings. This, so-called, N-duality was first discussed in
\cite{HH} in the context of four dimensional SYM.

However, we would like to stress that we show the same U-duality
maps the scaling limit of the backgrounds in one theory to the
scaling limit of the backgrounds in the other. On one hand the
decoupling limit gives that all the closed string and massive open
string modes are decoupled and thus the theory reduces to ordinary
Yang-Mills type theory. On the other hand the scaling limit of the
background, which are connected by U-duality from the former,
results in the decoupling of bulk closed string modes, and gives
the theory of open strings only in the background of
non-commutative space-time, thus gives NCOS theory. This tells
that the dual theory of $U(N)$ SYM with $M$ unit flux of electric
field is not $U(M)$ SYM with $N$ electric flux, but rather the
full NCOS theory on $M$ D-branes with $N$ unit flux of electric
field.

In the case of $T^1$ and $T^2$ compactifications, the duality maps
strong coupling regime of one theory to the weak coupling regime
in the other and vice versa \cite{GMSS}. In the case of $T^3$
compactification, the Yang-Mills coupling in matrix theory is
independent of the open string coupling in the NCOS theory. For
higher dimensional torus compactifications, the weak coupling
regime of one theory maps to the weak coupling regime of the
other. From these dualities, we can easily see that the Higgsing
process in matrix theory corresponds to extracting fundamental
string in the NCOS theory. This is easily confirmed by computing
binding energy which should be identical in both cases.

We also consider some bound states which have finite energy. Once
again, under the U-duality, bound states in one theory map to the
bound states in the other with the same finite energy, which
support the U-duality between those theories. In the case of
matrix theory on $T^4$ and $T^5$ compactifications, these bound
states play essential roles for the complete description of the
theory.

In section 2 we shortly review the appropriate scaling limit for
each class of theories. In section 3 we explain the U-duality
which connects the NCOS theory and the matrix theory for each
$T^p$ compactifications. In section 4 we draw some conclusions.

\section{Decoupling limit and $T^p$ compactifications}

In this section we review the scaling limit taken to get the NCOS
theory and the matrix theory, under which all the bulk degrees of
freedom are decoupled. We also discuss the torus compactifications
of those theories.

\subsection{ NCOS theory}

Consider a D$p$-brane with a near critical electric field in the
0,1 direction.
 The electric field on the brane is given by
\be
2 \pi \a'\e^{01}F_{01}=1-{\e\over 2}~,
\ee
where critical electric field corresponds to $\e=0$:
\be
\e^{01}F_{01}^{crit}={1 \over 2\pi \a'}~.
\ee
In the presence of background electric field, or equivalently
$B_{01}$ field on the D-brane worldvolume, the metric seen by bulk
closed string modes are different from the effective metric seen
by the open strings on the D-brane worldvolume. The background
metric for closed string is chosen to be \cite{GMSS}
\footnote{$\m, \n=0,1$ denote the electric directions on the brane
and $i,j=2,3, \cdots , p$ denote the remaining directions of the
brane. $M,N=p+1, \cdots , 9$ denote the directions transverse to
the brane. }
\be
g_{\m\n}= \eta_{\m\n}~, \ \ \ g_{ij} = \e  \d_{ij}~, \ \ \
g_{MN}=\e \d_{MN}~.
\label{met1}
\ee
 The effective string tension of the open string stretched in the
$x^1$ direction is
\be
{1\over 4 \pi \a'_e} \equiv {1 \over 2 \pi\a'} \left( 1 - 2\pi \a'
\e^{01} F_{01}\right)={\e \over 4\pi \a'}~.
\label{tension}
\ee
Therefore, while $\a'\equiv M_s^{-2}$ sets the scale of closed
string modes, $\a'_e\equiv M_e^{-2}$ can be considered as the
scale of open string modes stretched in the electric direction.
The effective open string metric, noncommutativity parameter and
string coupling $G_o^2$ corresponding to (\ref{met1}) can be
determined as \cite{FT,CLNY,SW}
\be
G_{\m\n}=\e\eta_{\m\n}~, \ \ \  G_{ij}=\e \d_{ij}~, \ \ \
\Th^{\m\n} =2\pi\a'_e \e^{\m\n}~, \ \ \ G_o^2=g_s \e^{1\over 2}~.
\ee
The NCOS limit \cite{SST,GMMS}, under which the bulk closed string modes are
decoupled, is given by $\e  \rightarrow 0$ while taking $\a'_e$,
$G_o$ fixed. Therefore, under this limit, the scaling is
summarized as
\bea
g_{\m\n}   \sim  \CO(1)~,\ \ \ g_{ij} \sim \e~,\ \ \ g_{MN} \sim
\e~,\ \ \ g_s \sim \e^{-{1\over 2}}~,\ \ \ \a' \sim  \e~.
\label{nscale0}
\eea

Note that if we have compactification along the circle in the
non-electric direction of D-brane worldvolume, then the scaling
(\ref{nscale0}) implies the proper radius of the circle behaves as
$R_i\sim \e^{1\over 2}\rightarrow 0$. Because of the scaling
behavior (\ref{nscale0}), the scaling of closed string metric,
string coupling and string tension after the T-duality is the same
as before the T-duality and thus again given by (\ref{nscale0}).
Therefore ($p+1$)-dimensional NCOS theory on $S^1$ in the
non-electric direction on the worldvolume is T-dual to a
$p$-dimensional NCOS theory on dual torus in the transverse
direction.

\subsection{Matrix theory on $T^p$}
Consider the system with $N$ D0-branes. We want to take the
scaling limit \cite{DKPS} under which the low energy description
is solely given by the degrees of freedom on $N$ D0-branes. To
decouple the bulk closed string modes we take the scaling limit
$\a' \rightarrow 0$ while keeping the energy of stretched open
strings between D0-branes with transverse distance $r$,
\[
E={r\over \a'}~,
\]
fixed. We also keep Yang-Mills coupling
constant $g_{YM}^2={g_s\over 2\pi (\a')^{3\over 2}}$ fixed. These
requirements fix the scaling of string coupling and the closed
string metric in terms of the scaling of $\a'$. Let
\be
\a'\sim \e^{1\over 2}
\ee
in the scaling limit $\e\rightarrow 0$. Then we have
\bea
g_s \sim \e^{3\over 4}~,\ \ \ g_{MN} \sim \e~,\ \ \ M_p \sim
\e^{-{1\over 2}}~, \ \ \ R   \sim \e~,
\label{mscale0}
\eea
where $M_p={M_s\over g_s^{1\over 3}}$ and $R={g_s\over M_s}$
denote eleven-dimensional Planck scale and eleventh circle radius,
respectively. In the $\e\rightarrow 0$ limit, we arrive at the
theory in which all the bulk closed string modes as well as the
massive modes of open strings\footnote{Note that, in this setup,
the scale of open string modes is the same as the scale of closed
string modes, $M_s$. It is also true in the sector with electric
flux.} attached between D0-branes are decoupled. Therefore the
remaining relevant degrees of freedom are massless modes of open
strings on the worldvolume of D0-branes and the system is
described by (0+1) dimensional SYM quantum mechanics. One may take
the discrete light-cone \cite{susskind} to lift to
eleven-dimensions and then by taking the large $N$ limit, we would
get the matrix model originally proposed in \cite{BFSS}.

One can get similar scaling limit for the system with $N$
D$p$-branes ($p\leq 3$) which keeps SYM degrees of freedom only.
It is given by
\bea
\a'\sim \e^{1\over 2}~,\ \ \ g_s \sim \e^{3-p\over 4}~,\ \ \
g_{MN} \sim \e~,
\label{mscale}
\eea
while the metric components in the longitudinal direction are kept
fixed in the limit. Therefore, for $p \leq 3 $, the system is
described by SYM theory with Yang-Mills coupling constant,
\be
g_{YM}^2= (2\pi)^{p-2}g_s (\a')^{p-3\over 2}~.
\label{gym}
\ee
For $p\geq 4$, the SYM theory is not renormalizable and typically
new degrees of freedom enter \cite{bound,strom,LST}. In this paper
we call all these worldvolume theories of D$p$-branes, M2-brane,
M5-brane and NS5-branes in the scaling limit
(\ref{mscale})\footnote{The scaling limit for NS5-brane is the
S-dual version of (\ref{mscale}).} as matrix theory.

Now consider the theory defined on $p$-torus starting from
(0+1)-dimensional matrix theory set-up \cite{seiberg,sen}. In this
case the proper torus radii $R_i\sim \e^{1\over 2}$ from
(\ref{mscale0}) are vanishingly small, thus it is natural to
perform T-duality along the $p$-torus. After T-duality, $N$
D0-branes turn into $N$ D$p$-branes wrapped in the dual torus with
radii
\[
\S_i = {\a' \over R_i} \sim  \CO(1)~.
\]
The scaling behavior of string coupling and string tension becomes
\bea
g'_s = g_s \prod_{i=1}^p{\sqrt{\a'}\over R_i}\sim \e^{3-p\over
4}~, \ \ \ \a' \sim \e^{1\over 2}~,\ \ \ g_{MN} \sim \e~,
\label{mscalep}
\eea
which is the same as (\ref{mscale}).

\section{U-duality of the NCOS theory on $T^p$}
In this section we describe the U-duality transformations of the
NCOS theories on $T^p$. The basic set-up is $N$ D0-branes with $M$
KK momentum along compactified $x^1$ in the scaling limit
(\ref{mscale0}). As explained in the previous section, after
T-duality, it becomes (1+1)-dimensional SYM theory. In matrix
theory all BPS states in M theory are realized as bound states in
SYM or its completions. The physics of M theory in the background
of those bound states would be described by the local fluctuations
on the corresponding bound states in matrix theory. In the case at
hand, the KK momentum turns into fundamental string wrapping
$S^1$. Therefore it becomes the bound state of $(N,M)$ strings
which is described by (1+1)-dimensional SYM theory in the sector
of $M$ units of electric flux. If we have further torus
compactifications, we simply take T-duality on those tori,
following the prescriptions given in the last section, and get the
matrix theory on $T^p$ with $M$ units of electric flux.

Another way to describe this is to take $x^1$ - $x^{11}$ flip.
This turns out to correspond to the NCOS theory. For $T^p$ with
$p=1,2$, this reduces to the known results \cite{GMSS}, namely,
the S-duality between NCOS theory and SYM. For $T^p$ with $p\geq
3$, we encounter new dual relations.

\subsection{(1+1)-dimensional NCOS theory}

We begin with $N$ sector of matrix theory on $S^1$ with background
$M$ KK momentum along the circle, in the scaling limit
(\ref{mscale0}). The T-duality along the circle, turns the system
into the theory of ($N$,$M$)-strings which is (1+1)-dimensional
$U(N)$ SYM theory with $M$ units flux of electric field. The
scaling of the background goes as
\be
g_{\m\n}=\eta_{\m\n}~, \ \ \  g_{MN}=\e \d_{MN}~, \ \ \ g_s\sim
\e^{1\over 2}, \ \ \ \a'\sim \e^{1\over 2}~.
\label{mscale1}
\ee

The allowed values of the electric field on the D-string are
quantized, and the quantum number can be interpreted as the number
of fundamental strings immersed in D-strings. The $M$ unit of
electric flux in $N$ D-strings can be expressed as
\bea
N{2 \pi \a'\e^{01}F_{01}\over \sqrt{1-(2\pi \a')^2 F^2}}=g_s M~.
\label{e-field}
\eea
Note that, if we take the scaling in matrix theory,
(\ref{mscale1}), the effective string tension of open strings
stretched in $x^1$ direction is the same as the string tension of
closed strings in bulk, $(2\pi\a')^{-1}$, as
$2\pi\a'\e^{01}F_{01} \rightarrow 0 $, independent of finite $N$
and $M$. Therefore we can conclude that, in the scaling limit
(\ref{mscale1}), it is described by (1+1)-dimensional ordinary
$U(N)$ SYM in the sector of $M$ units of electric flux.

Now we take different path, in which we consider $x^1$ as eleventh
circle. This amounts to the $x^1$ - $x^{11}$ flip. After
relabeling the coordinates, the scaling (\ref{mscale0}) maps to
\bea
M_p \sim \e^{-{1\over 2}}~, \ \ \ R   \sim  \e^{1\over 2}~,\ \ \
R_1 \sim \e~,\ \ \ R_M \sim  \e^{1\over 2}~.
\label{nscale}
\eea
Now after
compactification on this new $x^{11}$ circle, the scaling of
parameters in ten-dimensional IIA string theory becomes
\bea
g_s \sim \CO(1)~,\ \ \ \a' \sim \e~.
\eea
After T-duality along $x^1$, whose radius is vanishingly small,
the scaling goes as
\bea
R_1 \sim \CO(1)~,\ \ \ R_M \sim \e^{1\over 2}~,\ \ \ g_s \sim
\e^{-{1\over 2}}~,\ \ \ \a' \sim \e~,
\label{nscale1}
\eea
which is nothing but the scaling behavior of (1+1)-dimensional
NCOS theory (\ref{nscale0}). Indeed, under these chain of
dualities, $N$ D0-branes with $M$ KK momentum become $M$ D-strings
with background $N$ units of electric flux, which is given by
\be
2 \pi \a'\e^{01}F_{01}=1-{\e\over 2}={g_s {N\over M} \over
\sqrt{1+(g_s{N\over M})^2}}~.
\ee
Therefore $\e$, which represents the deviation of electric flux
from critical value, is given by
\bea
\e=\left({M\over N g_s}\right)^2~,
\eea
from which we find
\be
G_o^2={M\over N}~.
\label{Go}
\ee
 Therefore on the one end of the chain of
dualities, we have the (1+1)-dimensional NCOS theory of $M$
D-strings with $N$ units of near critical electric flux and on the
other end we have $U(N)$ SYM in a sector of $M$ units of electric
flux background, as summarized in table 1.

\begin{center}
\begin{tabular}{|c|c c c c c|} \hline
 Theory  & $g_s$  & $\a'$ & $x^1$ radius & D-string & E-flux \\
 %      &   &   & circumference & circumference  \\
 \hline
  NCOS  & $  \e^{-{1\over 2}}G_o^2 $
       & $  \e\a'_e $
       & $ r_1 $ & M & N \\ \hline
  SYM  & $ \e^{1\over 2}G_o^{-2} $
       & $ \e^{1\over 2}G_o^2 \a'_e$
        & $ r_1 $ & N & M \\ \hline
\end{tabular}
\\
[.3cm]
 Table 1. 1+1 dimensional theories
\end{center}

The Higgsing energy is identical in both theories \cite{GMSS,KM},
as a simple consequence of U-duality. In the SYM theory, the bound
state energy of $N$ D-string with $M$ units of electric flux
wrapping $x^1$-circle is given by
\be
H_{(M,N)}=\sqrt{\left({Nr_1\over g_s\a'}\right)^2+\left({Mr_1\over
\a'}\right)^2}~,
\label{energyh}
\ee
in which the first term in the square root divergies much faster
than the second one, under the scaling limit (\ref{mscale1}). The
energy of the ground state in $U(N)$ SYM with $M$ units of
electric flux above the $U(N)$ SYM ground state with no flux is
given by
\be
E_{(M,N)}=H_{(M,N)}-H_{(0,N)}=2\pi r_1{g_{YM}^2\over 2} {M^2\over
N} ~,
\ee
which is finite under the scaling limit (\ref{mscale1}). Here the
gauge coupling of $U(N)$ SYM theory is given by
\be
g_{YM}^2 = {g_s\over 2\pi \a'} = {1\over 2\pi G_o^4\a'_e}~,
\ee
from the table 1. Therefore, using (\ref{Go}) the ground state
energy is given by
\be
E_{(M,N)}= 2\pi r_1{N\over 4\pi\a'_e}~,
\label{energy-g}
\ee
and thus the energy needed to extract D-string is
\be
E= 2\pi r_1{1\over 4\pi\a'_e}~,
\ee
which is the same energy needed to separate an open string with
the effective tension $(4\pi \a'_e)^{-1}$ in (1+1)-dimensional
NCOS theory on $S^1$ with the radius $r_1$ \cite{GMSS,KM}.

One may understand this identification in the following way.
Consider the bound state energy of $M$ D-string with $N$ units of
electric flux wrapping $x^1$-circle, which is again given by
(\ref{energyh}) with interchanged $N$ and $M$. Now we would like
to impose the scaling limit (\ref{nscale1}) of NCOS theory. Then
the second term in the square root divergies much faster than the
first term. The ground state energy above the ground state energy
of $N$ fundamental strings only is given by
\be
E_{(N,M)}=H_{(N,M)}-H_{(N,0)}=2\pi r_1 N_1{M^2 \over  4\pi N^2
g_s^2\a'} = 2\pi r_1{N \over 4\pi\a'_e} ~.
\ee
Therefore we may consider the effective tension of the open string
in the NCOS theory as $(4\pi \a'_e)^{-1}$ which agrees with
(\ref{tension}). Of course, this is a simple consequence of
U-duality\footnote{The diverging energy, which is subtracted, is
also identical in both theories as, again, a consequence of
U-duality.}. As will be shown in the following subsections this
holds true in the higher-dimensional torus compactifications.

We also note that this dual relation holds even after taking the
decompactification limit $r_1\rightarrow \infty$. This can be
confirmed by noting that one can also get the same conclusion by
simply taking S-duality \cite{GMSS} starting from
(1+1)-dimensional NCOS theory.

\subsection{(2+1)-dimensional NCOS theory}
The dual relation between NCOS theory on $M$ D-branes with $N$
electric flux and $U(N)$ SYM with $M$ electric flux
 holds in higher dimensional $T^p$ compactifications. On $T^2$
compactifications, we find the dual theory of NCOS theory on $M$
D2-branes  with $N$ electric flux is $U(N)$ SYM theory in the
sector of $M$ units of electric flux. This is achieved by taking
another T-duality starting from the configurations in the previous
subsection. It is easily seen by examining the corresponding
scaling limit they require. In the matrix theory side, they are
given by (\ref{mscalep}) for $p=2$, while in the NCOS theory side,
after taking T-duality the scaling is given by
\bea
R_1 \sim \CO(1)~,\ \ \ R_i \sim \e^{1\over 2}~,\ \ \ R_M \sim
\e^{1\over 2}~,\ \ \ g_s \sim \e^{-{1\over 2}}~,\ \ \ M_s \sim
\e^{-{1\over 2}}~,
\label{nscale2}
\eea
which is the same scaling as (\ref{nscale0}). The relation between
these two theories are summarized in table 2, in which
$g_o^2\equiv G_o^{-1}(r_2M_e)^{3\over 2} $.

\begin{center}
\begin{tabular}{|c|c c c c c c|} \hline
 Theory  & $g_s$  & $\a'$ & $x^1$ radius &$x^2$ radius &
         $x^{11}$ radius & $M_p^3$ \\
 %      &   &   & circumference & circumference  \\
 \hline
  NCOS  & $  \e^{-{1\over 2}}G_o^2 $
       & $  \e\a'_e $ & $r_1$
       & $ \e^{1\over 2}r_2 $
       & $G_o^2\a_e^{\prime {1\over 2}}$
       & $\e^{-1}{M_e^3\over G_o^2}$ \\ \hline
  SYM  & $\e^{1\over 4}g_o^2 $
       & $ \e^{1\over 2}\left({G_o\over g_o}\right)^{4\over 3}\a'_e$
       &$r_1$
       & $G_o^2\a_e^{\prime {1\over 2}}$
       & $\e^{1\over 2}r_2 $
       & $\e^{-1}{M_e^3\over G_o^2}$
       \\ \hline
\end{tabular}
\\
[.3cm]
 Table 2. 2+1 dimensional theories
\end{center}

One can see that these two theories are S-dual each other. The
strong coupling regime of NCOS theory, $G_o^2\rightarrow \infty$,
maps to the weak coupling limit of SYM theory with gauge coupling,
\be
g_{YM}^2={g_s \over \sqrt{\a'}}=G_o^{-2}{r_2^2 \over
(\a'_e)^{3\over 2}}~.
\label{ymc2}
\ee
Viewing as T-dual version of (1+1)-dimensional NCOS theory, the
(2+1)-dimensional NCOS theory has open string coupling
\be
G_o^2={M\over N}(r_2 M_e)~.
\label{Go2}
\ee

In the $U(N)$ SYM theory with $M$ units of electric flux, the
ground state energy above the $U(N)$ SYM ground state without
electric flux is given by
\bea
E &=& \sqrt{\left({Nr_1l_2\over 2\pi g_s(\a')^{3\over 2}
      }\right)^2 +\left({Mr_1\over \a'}\right)^2}
      -{Nr_1l_2\over 2\pi g_s(\a')^{3\over 2} }\nonumber \\
  &=& 2\pi r_1 { g_{YM}^2\over 2 l_2} {M^2 \over N}~,
\eea
where $l_2= 2\pi G_o^2\a_e^{\prime {1\over 2}}$ is the proper size
of the $x^2$-circle in SYM theory. Therefore using (\ref{ymc2})
and (\ref{Go2}) the ground state energy is given by
\be
E = 2\pi r_1{N\over 4\pi\a'_e}
\ee
Therefore the energy cost needed for taking single D2-brane to
infinity is $ 2\pi r_1{1\over 4\pi \a'_e}$, the same energy cost
to liberate a fundamental string in (2+1)-dimensional NCOS theory.
Again this is a consequence of U-duality.

The above dual relation can be derived by directly interchanging
$x^2$ and $x^{11}$ coordinates in the (2+1)-dimensional NCOS
theory on the torus, as can be readily seen in the table 2.

\subsection{(3+1)-dimensional NCOS theory}
In this section we discuss the case of $p=3$. Here we have new
dual theories of (3+1)-dimensional NCOS theory. If $x^3$ is also a
compact coordinate, we can start from the above (2+1)-dimensional
theories, and take T-duality. In this way we find the dual theory
of NCOS theory on $M$ D3-branes wrapping on $ T^3$ with $N$ units
of electric flux are $U(N)$ SYM theory on $ T^3$ in a sector with
$M$ units of electric flux.  The (3+1)-dimensional NCOS theory has
open string coupling
\be
G_o^2={M\over N}{r_2 r_3\over \a'_e}~,
\label{Go3}
\ee
which is given by  T-dualizing (2+1)-dimensional NCOS theory on
the torus. The coupling constant of this SYM theory is given by
\be
g_{YM}^2=2\pi{r_2 r_3 \over \a'_e}~,
\label{ymc3}
\ee
where $r_2$, $r_3$ are defined in the table 3. The coupling is
independent of the original open string coupling constant $G_o$!
It is not strong-weak coupling duality as in the previous examples
and thus one may be able to compare the behavior of one weakly
coupled theory to another weakly coupled theory.

Since we are on IIB side, the theory has S-duality. By applying
\cite{GMMS} S-duality on the NCOS theory of $M$ D3-branes with $N$
units of electric flux, one obtains NCYM theory of $M$ D3-branes
with $N$ units of magnetic flux \cite{SW}. The magnetic field is
given by \cite{GMMS}
\bea
F_{23}={1\over \theta}
     &=&{\sqrt{-g}\over g_s} {\e^{01}F_{01}\over
         \sqrt{1-(2\pi\a')^2F_{01}^2}} \nonumber \\
     &=& {1\over 2\pi G_o^2\a'_e}~.
\eea
Thus the dimensionless parameter $\Theta = {1\over 2\pi r_2r_3}
\theta$ defined in \cite{SW} has rational value $\Theta ={M\over
N}$ after using (\ref{Go3}). In this case it was shown in
\cite{SW} that there is a T-duality which maps NCYM theory to
ordinary SYM theory.

By applying S-duality on the ordinary $U(N)$ SYM theory in the
sector of $M$ units of electric flux, one gets ordinary $U(N)$ SYM
theory in the sector of $M$ units of magnetic flux with gauge
coupling
\be
(g'_{YM})^2= {4\pi^2\over g_{YM}^2}~.
\label{ymc3p}
\ee
In fact this $U(N)$ SYM theory with $M$ units of magnetic flux is
the one mentioned above as a T-dual theory to the NCYM theory.

All these are summarized
in the table 3 where $g_o^2\equiv {g_{YM}^2\over 2\pi}={r_2 r_3
\over \a'_e}$.

\begin{center}
\begin{tabular}{|c|c c c c c|} \hline
 Theory  & $g_s$  & $\a'$ & $x^1$ radius  &
         $x^i$ radius & $g_{MN}$\\
 %      &   &   & circumference & circumference  \\
 \hline
  NCOS  & $  \e^{-{1\over 2}}G_o^2 $
       & $  \e\a'_e $
       & $ r_1 $
       & $ \e^{1\over 2}r_i $ & $\e\d_{MN}$ \\ \hline
  NCYM  & $ \e^{1\over 2}G_o^{-2} $
       & $ \e^{1\over 2}G_o^2 \a'_e$
        & $ r_1 $
       & $ \e^{1\over 2}r_i $ & $\e\d_{MN}$  \\ \hline
  SYM with&&&&&\\
  E-flux & $ g_o^2$
       & $ \e^{1\over 2} \left({G_o\over g_o}\right)^2 \a'_e $
       & $  r_1 $
       & $ \left({G_o\over g_o}\right)^2 r_i$& $\e\d_{MN}$
       \\ \hline
  SYM with &&&&&\\
  M-flux  & $ g_o^{-2}$
       & $  \e^{1\over 2} G_o^2 \a'_e $
       & $r_1 $
       & $\left({G_o\over g_o}\right)^2 r_i$& $\e\d_{MN}$ \\ \hline
\end{tabular}
\\
[.3cm]
 Table 3. 3+1 dimensional theories
\end{center}

Here again, the Higgsing energy in SYM theory corresponds to the
energy required to extract a string in NCOS theory. Since the case
with electric flux was presented in the previous sections for
lower dimensional D-branes, here we present only the case with
magnetic flux. In the $U(N)$ SYM theory with $M$ units of magnetic
flux, the ground state energy above the $U(N)$ SYM ground state
without magnetic flux is given by
\bea
E &=& \sqrt{\left({Nr_1l_2l_3\over g_s(2\pi\a')^2 }\right)^2
      +\left({Mr_1\over g_s\a'}\right)^2}
      -{Nr_1l_2l_3\over  g_s(2\pi\a')^2 }\nonumber \\
  &=& 2\pi r_1 { 2\pi^2\over(g'_{YM})^2 l_2l_3} {M^2 \over N}~,
\eea
where $l_i= 2\pi {G_o\over g_o}^2 r_i$ is the proper size of the
$x^i$-circle in SYM theory. Therefore using (\ref{Go3}),
(\ref{ymc3}) and (\ref{ymc3p}), the ground state energy is given
by
\be
E = 2\pi r_1{N\over 4\pi\a'_e}~,
\ee
And therefore the Higgsing energy in $U(N)$ SYM with $M$ units of
magnetic flux coincides with the energy cost to send a wrapping
string along $x^1$-direction to infinity.

\subsection{OM theory}
Next we consider the $T^4$ compactification. In the matrix theory
side, after taking T-duality along the torus, the string coupling
divergies, as can be seen from (\ref{mscalep}) with $p=4$.
Therefore we are considering the strong coupling limit of IIA
string theory, which is an eleven dimensional M theory. D4-branes
become M5-branes wrapping the eleventh dimension whose radius is
now finite and is given by
\bea
R_{11} =  g_s \sqrt{\a'} \sim \CO(1).
\label{rel6}
\eea

The low energy (4+1)-dimensional SYM theory on D4-brane
worldvolume is not renormalizable and breaks down at energies of
order $E\sim {1\over g_{YM}^2}$, where new degrees of freedom
should be taken into account. Since the SYM coupling is given by
$
g_{YM}^2=(2\pi)^2 R_{11}~,
$
the new degrees of freedom are the KK momentum in eleventh
direction, i.e. D0 particles in ten dimensional point of view.

The eleven dimensional Planck scale becomes
 \bea
 M_P
=\frac{ M_s}{( g_s)^{1/3}} \sim \e^{-1/6}\longrightarrow \infty~.
\label{rel7}
\eea
Therefore the bulk degrees of freedom are decoupled and it becomes
the theory of $N$ M5 branes in eleven dimensions, i.e. (0,2)
theory \cite{bound,strom},
with $M$ unit flux of background worldvolume three form
field strength.

In the dual NCOS theory, after taking T-duality along $T^4$, the
string coupling diverges, $g_s\sim \e^{-{1\over 2}}$, so we may
lift to the eleven dimensions again. Here the radius of eleventh
direction is fixed under the limit and in fact is the same as the
eleventh radius in the matrix theory above. The eleven dimensional
Planck scale becomes
$
M_P  \sim \e^{-1/3} \rightarrow \infty~.
$
Therefore (4+1)-dimensional NCOS theory can be considered as the
compactification of (5+1)-dimensional OM theory\cite{GMSS} with
finite radius. Indeed it gives the right scaling for the OM theory
with a background worldvolume three form field strength
\be
H_{012}= {2 \pi F_{01}\over R_{11}}=\e^{-1}2 M_{\rm eff}^3
\left(1-{\e\over 2}\right)~,
\ee
where $R_{11}=G_o^2(\a'_e)^{1\over 2}$ is the radius of the
eleventh coordinate. ${M_{\rm eff}^3\over 4\pi^2}\equiv
{M_e^3\over 8\pi^2G_o^2 }$ is the effective tension of a open
membrane  stretched in the $x^1$ and $x^{11}$ directions
\cite{GMSS}. This can be confirmed in the same way as the case of
effective string tension in section 3.1. One can  consider the
bound state energy of D4-branes with electric flux in the NCOS
scaling limit (\ref{nscale0}) and then reexpress in the eleven
dimensional variables. The relation between the (4+1)-dimensional
NCOS theory on $M$ D4-branes and matrix theory with $M$ units of
electric flux are summarized in table 4 with $g_o^2\equiv
G_o(r_2r_3r_4M_e^3)^{1\over 2} $.

\begin{center}
\begin{tabular}{|c|c c c c c|} \hline
 Theory  & $g_s$  & $\a'$ & $x^i$ radius &
         $x^{11}$ radius & $M_p^3$ \\
 %      &   &   & circumference & circumference  \\
 \hline
  NCOS  & $  \e^{-{1\over 2}}G_o^2 $
       & $  \e\a'_e $
       & $ \e^{1\over 2}r_i $
       & $G_o^2\a_e^{\prime {1\over 2}}$
       & $\e^{-1}{M_e^3\over G_o^2}$ \\ \hline
  matrix theory&&&&&\\
  with E-flux
       & $\e^{-{1\over 4}}g_o^2 $
       & $ \e^{1\over 2}\left({G_o\over g_o}\right)^4\a'_e$
       & $  \left({G_o\over g_o}\right)^4 r_i$
       & $G_o^2\a_e^{\prime {1\over 2}}$
       & $\e^{-{1\over 2}}\left({g_o\over G_o}\right)^4{M_e^3\over G_o^2}$
       \\ \hline
\end{tabular}
\\
[.3cm]
 Table 4. 4+1 dimensional theories
\end{center}

The NCOS theory on  torus has other degrees of freedom with finite
energy. These are D-branes wrapping on the non-electric
directions. In the (4+1)-dimensional NCOS theory, D0-branes and
D2-branes wrapping on 2-torus can have finite energy. Especially
D0-branes have the energy
\[
E={1\over R_{11}}={M_e\over G_o^2}
\]
thus in the strong coupling regime, $G_o > 1$, D0-brane excitation
has lower energy than the string stretched in the $x^1$ direction.
Under U-duality we are considering, the D0-brane in NCOS theory
maps to D0-brane in matrix theory on $T^4$. This identification
can be easily established as both theories have the same $X^{11}$
radius and thus D0-branes have the same excitation energies in
both theories.

\subsection{Near critical NS5-brane theory}
Finally we consider the compactification on $T^5$. In the matrix
theory side, after taking T-duality along $T^5$, (\ref{mscalep})
tells us that it becomes the strong coupling limit of D5-branes in
type IIB theory. Therefore we perform the S-duality and then it
becomes the theory of $N$  NS5-branes in type IIB theory with $M$
unit flux background. Note that the unit flux of electric field on
the worldvolume of D5-branes can be traded for the unit $B_{01}$
NS 2-form field by the gauge transformation, which turns into the
unit $C_{01}$ RR gauge field under S-duality. The theory has
vanishingly small string coupling, $ g_s\sim \e^{1\over 2}~, $ and
finite closed string tension in the limit $\e\rightarrow 0$. The
original gauge coupling (\ref{gym}) turns, under S-duality, into
$
g_{YM}^2=(2\pi)^3\alpha'~.
$
The low energy effective gauge theory on the NS5-branes is not
renormalizable and breaks down at energies of order $E\sim {1\over
g_{YM}}\sim {1\over \sqrt{\a'}}$ where new degrees of freedom
enter. They are strings with tension $(2\pi \a')^{-1}$ and thus
can be interpreted as the fundamental strings on NS5-branes. Since
$g_s\rightarrow 0$, the bulk degrees of freedom are decoupled and
those strings remain on the NS5-branes. This theory is known as
LST (little string theory) \cite{LST}. By taking T-duality, one
can get LST on type IIA NS5-branes.

In the NCOS side, after performing T-duality along $T^5$, the
theory becomes (5+1)-dimensional NCOS theory on D5-brane, where
the open fundamental string stretched in the electric direction
remains light degrees of freedom. In this case the theory is in
type IIB side and thus we can take S-duality. The resultant theory
is the decoupled theory on the worldvolume of the NS5-brane with a
near critical background $C_{01}$ RR gauge field. In this case the
light degrees of freedom are open D-strings due to a near critical
two form RR potential and the theory is called OD1 theory
\cite{GMSS} or (1,1) OBLST \cite{Harm2}. These are S-dual version
of fundamental strings with a near critical two form NS-NS
potential, or electric field on the original D5-brane. When the
OD1 theory is defined on the torus, as in the present case, we can
take T-duality and the theory becomes OD$p$ theory on the
NS5-brane worldvolume, whose light degrees of freedom are open
D$p$-branes stretched in the direction of RR $C_{p+1}$ fields.
Note that for even $p$ the theory is the worldvolume theory of
type IIA NS5-brane and for odd $p$ it is the worldvolume theory of
type IIB NS5-brane, just like in the matrix theory.  In table 5,
the relations between these theories are summarized.  Here we show
the case of OD1 theory only. Other OD$p$ theories for $p\neq 1$
can be easily given by taking appropriate T-duality.

\begin{center}
\begin{tabular}{|c|c c c|} \hline
 Theory  & $g_s$  & $\a'$ & $x^i$ radius \\
 %      &   &   & circumference & circumference  \\
 \hline
  NCOS theory  & $  \e^{-{1\over 2}}G_o^2 $
       & $  \e\a'_e $
       & $ \e^{1\over 2}r_i $ \\ \hline
  OD1 theory  & $  \e^{1\over 2}G_o^{-2} $
       & $  \e^{1\over 2} G_o^2\a'_e $
       & $ \e^{1\over 2}r_i $
       \\ \hline
  LST&&&\\
  with E-flux & $\e^{1\over 2}G_o^{-2} $
       & $  G_o^4 {\a_e^{\prime 3}\over r_2r_3r_4r_5} $
       & $ G_o^2{\a_e^{\prime 2}\over r_2r_3r_4r_5} r_i$
       \\ \hline
\end{tabular}
\\
[.3cm]
 Table 5. 5+1 dimensional theories
\end{center}

The (5+1)-dimensional NCOS theory and OD1 theory on the torus also
has additional degrees of freedom with finite energy. These are
related to the little strings in LST under U-duality. In the NCOS
theory they are D-strings wrapped in one toroidal direction. In
the OD1 theory they are fundamental strings wrapped in the same
direction \footnote{After T-duality the theory becomes the OD$p$
theory on the dual torus and these finite energy excitations
correspond to KK momentum in the T-dualized direction.}. By U-dual
chain, all these map to the little string wrapped in the same
direction in LST. For example, one can easily see from the table 5
that the energy of fundamental strings wrapped in the
$x^5$-coordinate in OD1 theory is given by
\[
E={r_5 \over G_o^2\a'_e}~,
\]
which is the same as the energy for the little string in LST and
D-string in NCOS theory, wrapped in the same direction. One can
also easily confirm that, along the same line of arguments shown
in the lower dimensional toroidal compactifications, all these
theories on torus have the same Higgsing energy due to U-duality.
The corresponding process in NCOS theory is the separation of open
fundamental string wrapped in the $x^1$ direction to the infinity.
In OD$p$ theory with $C_{0\cdots p}\neq 0$ it corresponds to the
separation of D$p$-brane wrapped in the $x^1, \cdots x^p$
directions to the infinity. Note that in this case U-duality maps
weakly coupled regime of OD1 theory to weakly coupled regime of
LST.

\section{Discussions}
Now there are two (or more) different ways to study the
worldvolume theory on $N$ D$p$-branes with $M$ units of electric
flux and their analogues in M5-branes and NS5-branes. One is to
take the scaling limit (\ref{mscalep}), under which all the bulk
degrees of freedom and massive modes of open strings attached to
the branes are decoupled and the theory reduces to $U(N)$ SYM
theory or its generalization for $p\geq 4$ in the sector of $M$
units of electric flux. Another way to study the system is to take
the scaling limit (\ref{nscale0}), under which the bulk closed
string modes are again decoupled, while the massive modes of open
strings stretched in the electric direction can not be neglected
and thus the theory becomes full open string theory defined on
noncommutative spacetime. What has been shown in this paper is
that these are connected by U-duality inherited from M theory,
with the exchange of the number of branes and the number of units
of electric flux, i.e. they lie in the same moduli space. We can
identify the bound states with finite energy in the NCOS theory
with the states in the matrix theory. Some of these states
correspond to the elementary excitations in matrix theory on $T^4$
and $T^5$ compactifications. These dual relations nicely explain
that NCOS theories reduces to matrix theory at low energies
\cite{GMSS}.

From the dual relations between NCOS theory and DLCQ M theory, one
would get lots of useful information and insights on those
theories. In particular, it is quite striking that we can study
the large $N$ limit of matrix theory on torus, in the sector of
$M$ unit of electric flux, using the NCOS theory.

This U-duality would have deep implications in the dual gravity
descriptions of those theories. We will return this issue in the near
future.

\section*{Acknowledgements}
This work was supported in part by grant No. 2000-1-11200-001-3
from the Basic Research Program of the Korea Science and
Engineering Foundation.

\end{document}